\documentclass[iop]{emulateapj}

\newcommand{\boldforreferee}{}

\slugcomment{}

\shorttitle{Up to 100\,000 reliable strong gravitational lenses}
\shortauthors{Serjeant}

\begin{document}


\title{Up to 100\,000 reliable strong gravitational lenses in future dark energy experiments}


\author{S. Serjeant}
\affil{Department of Physical Sciences, The Open University, Milton Keynes, MK7 6AA, UK}




\begin{abstract}
  The {\it Euclid} space telescope will observe $\sim10^5$ strong
  galaxy-galaxy gravitational lens events in its wide field imaging
  survey over around half the sky, but identifying the gravitational
  lenses from their observed morphologies requires solving the
  difficult problem of reliably separating the lensed sources from
  contaminant populations, such as tidal tails, as well as presenting
  challenges for spectroscopic follow-up redshift campaigns. Here I
  present alternative selection techniques for strong gravitational
  lenses in both {\it Euclid} and the Square Kilometer Array,
  exploiting the strong magnification bias present in the steep end of
  the H$\alpha$ luminosity function and H{\sc i} mass function. Around
  $10^3$ strong lensing events are detectable with this method in the
  {\it Euclid} wide survey. While only $\sim1\%$ of the total haul of
  {\it Euclid} lenses, this sample has $\sim100\%$ reliability, known
  source redshifts, high signal-to-noise and a magnification-based
  selection independent of assumptions of lens morphology. With the
  proposed Square Kilometer Array dark energy survey, the numbers of
  reliable strong gravitational lenses with source redshifts can reach
  $10^5$.
\end{abstract}





\section{Introduction}\label{sec:introduction}

The next generation of optical/near-infrared survey space telescopes,
such as the {\it Euclid} 15\,000 deg$^2$ wide survey and the {\it
  WFIRST} 2000 deg$^2$ high latitude survey (Laureijs et al. 2011,
Spergel et al. 2013), aim to make pioneering constraints on the
evolving equation of state of dark energy.  Their slitless H$\alpha$
spectroscopy surveys use Baryonic Acoustic Oscillations to provide a
cosmological standard ruler while the imaging surveys provide weak
lensing constraints on dark energy.  The full Square Kilometer Array
(SKA) has also been proposed to be used to make comparable constraints
on dark energy (e.g. Abdalla et al. 2010), through a hemispheric H{\sc
  i} redshift survey. In the process, every one of these dark energy
experiments will also create huge legacy data sets that will benefit
many areas of astronomy: e.g. $5\sigma$ spectroscopic depths of
$4.3\times10^{-16}$ ergs\,cm$^{-1}$\,s$^{-1}$ and $0.36\times10^{-16}$
ergs\,cm$^{-1}$\,s$^{-1}$ for {\it Euclid} and {\it WFIRST}
respectively, and $5\sigma$ point source depths of $Y=J=H=24$ for {\it
  Euclid} and $Y=H=26.7$, $J=26.9$ for {\it WFIRST}.

Strong gravitational lensing particularly benefits from {\it Euclid}'s
$1.2$m diffraction-limited optical/near-infrared imaging over around
half the sky, with $\sim10^5$ galaxy-galaxy lensing events predicted
in the 15\,000 deg$^2$ wide survey (Laureijs et al. 2011). This is many
orders of magnitude of improvement in the numbers of lenses. Many
applications of strong gravitational lensing are limited by sample
size (e.g. Treu 2010), such as dark matter halo substructure
(e.g. Dalal \& Kochanek 2002), halo density profiles (e.g. Gavazzi et
al. 2007, 2008), cosmological parameters (e.g. Short et al. 2012,
Gavazzi et al. 2008, but see Schneider 2014), and extreme high
magnification systems (e.g. Wang \& Turner 1996). 

However, identifying these systems with high reliability and
completeness is not without its difficulties. In a heroic effort,
Jackson (2008) visually examined 285\,423 galaxies with $I<25$ in the
Cosmic Evolution Survey (COSMOS). He found two certain new lenses, one
probable candidate and a further $112$ candidates, to add to the
then-known $20$ COSMOS lenses and $47$ candidates. False positives
could be due, for example, to chance lensing-like locations of H{\sc
  ii} regions in a galaxy, or tidal tails resembling lensed
arcs. Visual inspection of larger data sets is a much more challenging
proposition. This problem has stimulated both the development of
arc-finder algorithms in multi-wavelength imaging data (e.g. Gavazzi
et al. 2014) and the creation the {\it Spacewarps} mass participation
citizen science experiment\footnote{\tt http://spacewarps.org}.

In this paper I present alternative methods of identifying strong
gravitational lens events, by exploiting the strong magnification bias
in the high-luminosity end of the H$\alpha$ luminosity function and
the high-mass end of the H{\sc i} mass function.  This paper follows
the unfashionably conventional format of introduction, method,
results, discussion and conclusions in sections
\ref{sec:introduction}, \ref{sec:method}, \ref{sec:results},
\ref{sec:discussion} and \ref{sec:conclusions} respectively.  This
paper assumes a Hubble constant of
$H_0=70$\,km\,s$^{-1}$\,Mpc$^{-1}=100h$\,km\,s$^{-1}$\,Mpc$^{-1}$ and
density parameters of $\Omega_{\rm M}=0.3$ and $\Omega_\Lambda=0.7$
throughout.

\begin{figure*}
\centering
\vspace*{-3cm}
\plotone{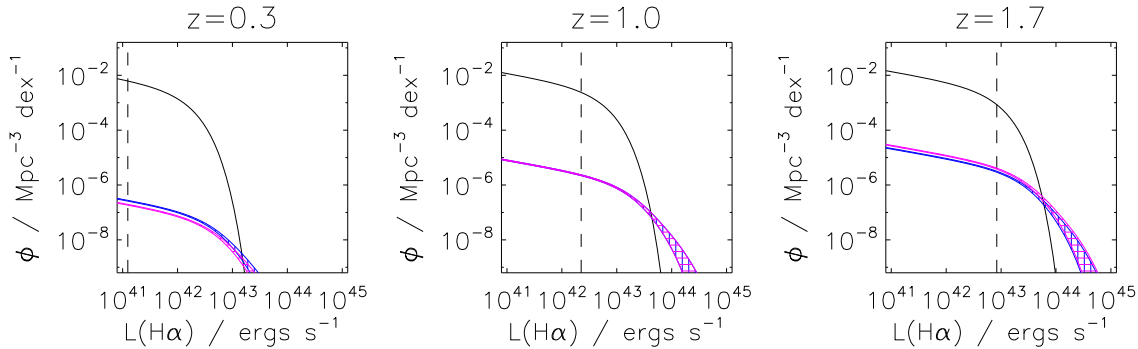}
\vspace*{-2.5cm}
\caption{\label{fig:halpha_lf} H$\alpha$ luminosity function from
  Geach et al. 2010 (black line) at redshifts $z=0.3$, $1.0$ and
  $1.7$, the singular isothermal sphere lenses of Perrotta et al. 2002
  and Perrotta et al. 2003 (pink), and with a non-evolving population
  of lenses normalised to the Perrotta predictions at an arbitrarily
  chosen $z=1$. The hatched regions span the range of maximum
  magnification $\mu\leq10$ and $\mu\leq30$. The vertical lines show
  the fiducial $5\sigma$ flux limit of the {\it Euclid} wide-field
  survey (Laureijs et al. 2011). Note
  that at $>12L_*$ the observed population is
  dominated entirely by strong gravitational lens systems.}
\end{figure*}

\begin{figure*}
\centering
\vspace*{-3cm}
\plotone{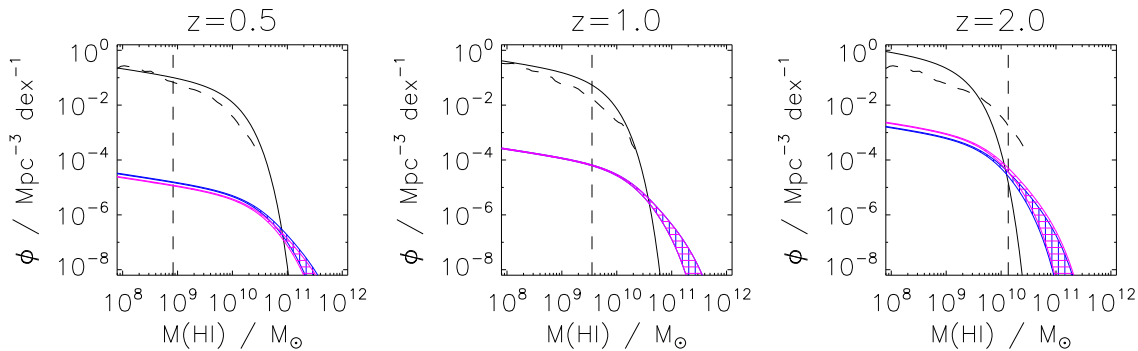}
\vspace*{-2.5cm}
\caption{\label{fig:hi_lf} H{\sc i} mass function following Abdalla et
  al. (2010) model ``C'' at redshifts of $z=0.5$, $1.0$ and $2.0$.
  {\boldforreferee Lensing models as in Fig.\,\ref{fig:halpha_lf}.}
  The vertical lines show
  the fiducial flux limit of the proposed full {\it SKA} dark energy
  survey from Abdalla et al. (2010), assuming a sensitivity parameter
  $f=1$ (see Abdalla et al. 2010 equation 2), $30$ km\,s$^{-1}$
  channels and a $100$\,deg$^2$ field of view. The dashed curve shows
  the evolving semi-analytic predictions of Lagos et al. (2011), for
  comparison. Similarly to Fig.\,\ref{fig:halpha_lf}, note that at
  $>12M_*$ the observed population is dominated entirely by strong
  gravitational lens systems.}
\end{figure*}

\section{Method}\label{sec:method}


In the presence of very steep number counts or luminosity functions, a
strong magnification bias can operate. The low probability of
galaxy-galaxy lensing producing an apparently-bright lensed galaxy can
be more than compensated for by the presence of a very large
population of faint candidate background lensed sources. If the slope
of the differential numbers steepens at progressively brighter fluxes
or luminosities, then the fraction of lensed sources also
progressively increases.

This method of selecting strong gravitational lenses has a successful
track record. Negrello et al. (2010) used this technique to select a
sample of strong lenses using simply a $500\,\mu$m flux limit of
$S_{500\,\mu\rm m}\geq100\,$mJy, exploiting the steep slope of submm
number counts. Half of the selection were contaminant (non-lensed)
local galaxies and blazars, but these were easily identifiable in
supplementary optical and radio survey data, leaving a sample with an
ostensible $\sim100\%$ reliability that the authors confirmed with a
wide range of spectroscopic and imaging follow-ups (see also
e.g. Viera et al. 2013, Bussmann et al. 2013). 
{\boldforreferee Gonz\'alez-Nuevo et al.} (2012)
extended this methodology to the high-luminosity end of the submm
luminosity function (estimated using submm photometric redshifts), and
in the process expanded the number of gravitational lens candidates in
{\it Herschel} surveys from $\sim100$ with a bright flux limit to
$\sim1000$ with a bright luminosity limit in the Herschel {\it ATLAS}
project (Eales et al. 2010). After removing known contaminants the
reliability of this strong gravitational lens sample was estimated to
be $\sim70\%$, sufficient to warrant extensive follow-up campaigns
(e.g. Amber et al. in prep.)  that could yield constraints on
cosmological parameters (e.g. Short et al. 2012), constraints on the
evolution of dark matter halo properties, and/or constraints on the
background lensed galaxy population (e.g. Eales 2014).

The H$\alpha$ and far-infrared luminosities of star-forming galaxies
have in common that they are dominated by the luminous output of
massive stars, which excites nebular emission in H{\sc ii} regions in
the case of H$\alpha$, and which is absorbed by dust and re-emitted as
grey-body radiation in the case of the far-infrared (e.g. Condon 1992
et seq.). Both H$\alpha$ and far-infrared luminosities are widely used
as star formation rate estimators, among other indicators. However,
the H$\alpha$ and far-infrared luminosities have different
dependencies on star formation rate: at high star formation rates, a
greater proportion of star formation is present in giant molecular
cloud complexes, so the H$\alpha$ luminosity will tend to plateau as
star formation rate increases while the far-infrared continues to
increase (e.g. Flores et al. 2004, Swinbank et al. 2004, Calzetti et
al. 2010). Balmer lines are inevitably dominated by low-extinction
regions (e.g. Serjeant et al. 2002), so even Balmer
decrement-corrected H$\alpha$ luminosities will under-predict the
total star formation rate in the most extreme starbursts. Unless the
extinction is accounted for empirically (e.g. Calzetti et al. 2010),
naive interpretations of H$\alpha$ luminosities consistently
under-predict star formation rates for the most luminous starbursts at
all redshifts (e.g. Sedgwick et al. 2013). {\boldforreferee For
  example, the H$\alpha$ luminosities of hyperluminous infrared galaxies
  ($>10^{13}L_\odot$) are not $\sim100\times$ brighter than
  those of luminous infrared galaxies ($>10^{11}L_\odot$). Therefore,
  extreme H$\alpha$-emitters should be rarer than extreme far-infrared
  emitters.}

For this reason, it is reasonable to expect the H$\alpha$ luminosity
function to have a bright-end slope that is steeper than its
far-infrared counterpart (e.g. Lapi et al. 2011, 
Marchetti et al. in prep.). The latter has already been
demonstrated to facilitate strong gravitational lens selection
(e.g. 
{\boldforreferee
Gonz\'alez-Nuevo et al. 2012}), so one would expect the bright end
of the H$\alpha$ luminosity function to be similarly useful. This
paper adopts the H$\alpha$ luminosity function determination of Geach
et al. (2010) at redshifts $0<z<2$, which is characterised by a
Schechter function. Note that this determination includes both
star-forming galaxies and active galaxies. 

Another population statistic where theoretical models and (to some
extent) observations lead to steep number counts and an expectation of
strong magnification bias is the neutral hydrogen galaxy mass
function. The physical processes determining the neutral hydrogen
distribution are arguably simpler than the physics of star formation
and dust obscuration.  The local H{\sc i} mass function is
well-described by a Schechter function (Zwaan et al. 2003), in
agreement with the expectations from semi-analytic models (e.g. Baugh
et al. 2004). At higher redshifts there are few empirical constraints,
though Abdalla et al. (2010) used damped Lyman $\alpha$ system number
densities and the evolving volume-averaged star formation history to
determine the evolution of the break in the mass function, resulting
in their preferred model, denoted model ``C''.  The semi-analytic
predictions of Lagos et al. (2011) evolve in the same sense as
Abdalla's model ``C'' but the evolution is less pronounced.
Therefore, in order to span the range of plausible predictions, this
paper adopts the model ``C'' of Abdalla et al. (2010) and a
conservative no-evolution model.

The formalism for calculating the differential magnification
probability distribution $p(\mu,z){\rm d}\mu$ for a magnification
$\mu$ is summarised in Blain (1996), Perrotta et al. (2002) and
Perrotta et al. (2003). Regardless of the lens population, the high
magnification tail has the form $p(\mu,z)=a(z)\mu^{-3}$ for some
function $a(z)$. For the purposes of this paper, ``strong'' lensing is
taken to mean magnifications $\mu\geq2$, in which this expression
applies. 

The form of $a(z)$ depends on the nature and evolving number density
of lenses. Both Blain (1996) and Perrotta et al. (2002, 2003) consider
several options for galaxy lens populations. 
In the case of Perrotta et al. (2002, 2003), the mass spectrum follows
the Sheth and Tormen (1999) formalism.  There is good evidence for the
galaxy lens population being well-described by singular isothermal
ellipsoids (e.g. Gavazzi et al. 2007, 2008) {\boldforreferee
as opposed to
Navarro Frenk White (1997) profiles, so this paper assumes a
singular isothermal sphere profiles} for the Perrotta et
al. lenses.  For an alternative approach, one can assume the mass
spectrum of lensing galaxies does not evolve with redshift (following
e.g. the approach of model ``B'' of Blain 1996, though with a
different cosmology), and normalise $a(z)$ to the prediction of
Perrotta et al. (2002, 2003) at an arbitrary redshift of $z=1$.  These
options should span the parameter space of plausible lens populations.

A more significant source of uncertainty in the lensing predictions is
the maximum magnification imposed by the finite source sizes. This
paper follows Perrotta et al. (2002) by spanning the range of
plausible maximum magnifications with $\mu\leq10$ and $\mu\leq30$, as
appropriate for source characteristic radii of $\sim1-10h^{-1}$\,kpc
(e.g. Wuyts et al. 2013).  High magnification events are known in
far-infrared and submm-selected lenses (e.g. Swinbank et al. 2010),
but this paper conservatively neglects this extreme population.  The
surface density of strong gravitational lenses increases with the
maximum magnification; the choice in this paper is more conservative
than that of e.g. Blain (1996) who used $\mu\leq40$, and Lima et
al. (2010) who used $\mu\leq100$.

\section{Results}\label{sec:results}
Fig.\,\ref{fig:halpha_lf} shows the effect of applying the strong
lensing formalism discussed above to the Geach et al. (2010) H$\alpha$
luminosity function. It is clear that at apparent luminosities
$>10^{44}$\,ergs\,s$^{-1}$, the observed population is entirely
dominated by strong gravitational lens systems, regardless of the
maximum magnification or the nature of the lensing population.  A
redshift-dependent threshold based on the shape of the evolving
unlensed luminosity function would yield a larger sample of lensed
galaxies.  This paper adopts $>12L_*$ as a conservative fiducial limit
for selecting a sample of strongly lensed H$\alpha$ emission line
galaxies. These samples have $\sim97-99\%$ reliablility, i.e. the
unlensed contaminants are $1-3$\% of the total sample selected in this
way. The fiducial H$\alpha$ $5\sigma$ flux limit for the {\it Euclid}
wide-area survey of $4.3\times10^{-16}$\,ergs\,s$^{-1}$\,cm$^{-2}$
(Laureijs et al. 2011) is indicated as a vertical line in
Fig.\,\ref{fig:halpha_lf}; the strong lensing systems will be detected
at very high signal-to-noise.

Assuming a fiducial sky coverage of $15\,000\,$deg$^2$ for the
wide-area {\it Euclid} mission survey, the numbers of strongly-lensed
galaxies with luminosities $>12L_*$ range from 1079 
using the no-evolution model and $\mu\leq10$ to 2717 with the
Perrotta et al. (2002, 2003) model and $\mu\leq30$. 

An even larger haul of lenses is possible with the SKA H{\sc i}
surveys.  With the H{\sc i} mass limit estimated by Abdalla et
al. (2010) for the fiducial one-year dark energy survey with the full
SKA, and assuming a survey area of $15\,000\,$deg$^2$, the $>12M_*$
limit yields of the order $10^4$ strong gravitational lenses with
$96-98$\% reliability. The
predictions range from 5800 lenses with the no-evolution lens
population and $\mu\leq10$ to 14\,000 with the Perrotta et al. (2002,
2003) model and $\mu\leq30$. The surface density of strong lenses is
higher than the {\it Euclid} H$\alpha$ case, because the mass function
Schechter normalisation is $\sim10\times$ higher than that that of the
H$\alpha$ Schechter function.

Model predictions of the evolution of the H{\sc i} mass function tend
to dismantle local high-mass systems into a larger number of
lower-mass systems at high redshift. This moves the characteristic
mass scale $M_*$ to lower masses at high redshifts, and the
normalisation $\phi_*$ to higher number densities. The effect on the
lensing predictions is to greatly increase the strength of the
magnification bias and the numbers of strong lensing systems. Figure
\ref{fig:hi_lf} shows the predictions for the evolving mass function
models. The SKA dark energy survey results in an astonishingly high
surface density of lenses with $97-99$\% reliability, of the order
$10^5$ in a $15\,000$\,deg$^{-2}$ survey: predictions range from 7400
lenses with the unevolving lens population and $\mu\leq10$, to
$190\,000$ lenses with $\mu\leq30$ and the Perotta models. 
{\boldforreferee 
The Lagos
et al. (2011) models are in fairly good agreement with the Abdalla et
al. (2010) predictions at $z<1$, but are discrepant at $z=2$
(Fig.\,\ref{fig:hi_lf}), though still following a Schechter shape. In
the Lagos case, the $>12M_*$ threshold will be brighter
than the Abdalla model, with high-redshift lensing
numbers intermediate between the Abdalla and no-evolution cases.}




\section{Discussion}\label{sec:discussion}
Are the H$\alpha$ emission lines in the lensing systems detectable by
{\it Euclid} against the glare of the foreground lens? A simple
calculation shows that this is typically not likely to be a
problem. The median $H$-band magnitude of the SLACS lenses is $H=15.9$
(Auger et al. 2009), equivalent to a flux {\it density} of
$4.7\times10^{-16}$\,W\,m$^{-2}$\,$\mu$m$^{-1}$. At $z=1.5$, the
$>12L_*$ selection is equivalent to a flux limit of
$5.7\times10^{-18}$\,W\,m$^{-2}$, approximately a factor of ten above
the point source flux limit (Laureijs et al. 2011). With the
reasonable assumption of an unresolved line at the expected resolving
power of $\lambda/\Delta\lambda=500$, the observed line flux {\it
  density} would be around four times the continuum level from the
lens. This is conservatively assuming the lensing galaxy is not
extended, which would dilute the flux density contribution from the
lens. This is also assuming that the H$\alpha$ emission is also
unresolved. The 3D-HST project has found effective H$\alpha$ radii at
$z\sim1$ of $\sim2-4$\,kpc (Wuyts et al. 2013), equivalent to
$0.25-0.5$ arcseconds, small compared to the NISP spectrometer pixel
scale of $0.5$ arcseconds. With angular magnification from lensing, we
may expect the H$\alpha$ emission to be typically no more than
marginally resolved by {\it Euclid}, apart from e.g. quad lens
configurations. 

The high surface density of strong galaxy-galaxy gravitational lens
events in the proposed one-year SKA weak lensing survey may have a
significant effect on the shear statistics. The strongly lensed
population is identifiable at high H{\sc i} masses, but further work
is needed to quantify the effect on the shear statistics of strongly
lensed galaxies at lower H{\sc i} masses. However, as a resource in
themselves, the H{\sc i}-selected strong lensing systems are
potentially revolutionary. {\boldforreferee There are no obvious
  significant unlensed contaminant populations for the H{\sc
    i}-selected samples, but dedicated lensing H{\sc i} surveys with
  {\it SKA} pathfinders/precursors would identify any such
  populations.}

Having prior knowledge of the background source redshifts partly
solves one of the major practical problems with strong gravitational
lens surveys: the multi-wavelength follow-ups. For example, most
COSMOS gravitational lens systems still lack a secure measurement of
the background redshift (e.g. Faure et al. 2011), including all three
lens candidates from Jackson (2008). This problem is intrinsic to
morphologically-selected samples of lensing systems and will be a
major challenge for morphologically-selected strong lenses from {\it
  Euclid} and {\it WFIRST}.

The H$\alpha$ and H{\sc i}-selected systems will nonetheless need
redshift estimation of the foreground lenses. The lensing optical
depth is maximised approximately half-way between the source and
observer, so lens redshifts will be typically in the range
$z\simeq0.5-1$. Optical and near-infrared ground-based photometric
surveys such as the Large Synoptic Survey Telescope will be
well-suited to obtaining photometric redshifts of massive galaxies at
these redshifts. Spectroscopic absorption line lens redshifts are
likely to require $\sim1$ hour exposures on $4$m-class optical
telescopes (e.g. Bussmann et al. 2013, Amber et al. {\it in prep}.).
For early-type lenses selected with {\it Euclid} and {\it WFIRST},
additional lens redshift estimation will be possible from the
fundamental plane (e.g. Serjeant et al. 1995, Eisenhardt et
al. 1996). The design of the Sloan Lens ACS Survey (SLACS) means it is
inevitably dominated by early-type lenses (e.g. Auger et al. 2009),
but it is not at all clear that early-type lenses are necessarily
responsible for nearly all the lensing optical depth in the
Universe. By exploiting magnification bias, the H$\alpha$ and H{\sc i}
selection methods are identifying high magnification lines of sight,
irrespective of the nature of the foreground lenses. Submm samples are
similarly selected on magnification bias, and as well as early-type
lenses they include several group lenses, cluster lenses and late-type
galaxy lenses (e.g. Negrello et al. 2010, Bussmann et al. 2013,
Negrello et al. 2014). The H$\alpha$ and H{\sc i} lensing samples will
be ideal to examine the diversity in the origin of the lensing optical
depth to the high redshift Universe.

Having a large sample of lensing systems will make a wide range of new
science applications of lensing possible. Rare lens configurations
will become available to study, such as extreme high magnification
events probing the morphologies of background galaxies at
$\sim100\times$ improved angular resolution or more (conservatively
excluded from our modelling in section \ref{sec:method}), or rare
``jackpot'' or ``double jackpot'' lenses with multiple galaxies at
different redshifts aligned along the line of sight (e.g. Gavazzi et
al. 2008). The latter systems can yield very useful constraints on
dark matter density profiles, by providing two or more lines of sight
past the closest galaxy. They were also originally envisaged as a
probe of cosmological parameters, but a generalisation of the
mass-sheet degeneracy has been discovered (Schneider 2014) that will
limit this application, unless additional modelling assumptions are
invoked.  Further statistical constraints on halo density profiles
will come from the number counts and redshift distributions of lenses
and sources (e.g. Short et al. 2012), and from the interface of strong
and weak gravitational lensing (e.g. Gavazzi et al. 2007).  Image flux
ratio anomalies have also been recognised as being caused by halo
substructure (e.g. Dalal \& Kochanek 2002, Vegetti et al. 2012), 
so large samples of lens
system will directly constrain semi-analytic models, particularly
using the {\it SKA} lensing samples that are immune to the effect of
dust obscuration in the foreground lenses. Large lens samples will
also make it possible to investigate the stellar initial mass function
as a function of galaxy types (e.g. Spiniello et al. 2011, Dutton et
al. 2013).

The background sources will at best only be marginally resolved by
their discovery observations with {\it Euclid}, {\it WFIRST} or {\it
  SKA}, but they will be among the brightest line emitters on the sky,
so will be easily amenable to follow-up observations. The high angular
resolution of $30-40$m-class telescopes will be very well suited to
high surface brightness arcs. The H$\alpha$-emitters will be
unchallenging targets in H$\alpha$ for the Thirty Metre Telescope
(TMT) or the European Extremely Large Telescope (E-ELT), and a 1-hour
exposure with e.g. the METIS instrument on E-ELT will be sufficient to
detect and resolve Pa$\alpha$ in the H$\alpha$-selected
sample. Scaling from local ultraluminous star-forming galaxies
(Veilleux, Kim \& Sanders 1999, Farrah et al. 2013), many
H$\alpha$-selected lensed galaxies will be easily detectable in
far-infrared lines with the proposed SPICA space observatory
(e.g. Roelfsema et al. 2012). Diagnostic line ratios may be subject to
differential magnification effects (e.g. Serjeant 2012), but this may
be corrected for in single systems or statistically in the population
through source and lens modelling. Alternatively, differential
magnification may be indirectly addressed through a comparison of
emission line profiles with those of spatially resolved lines
(e.g. Omont et al. 2013).

\section{Conclusions}\label{sec:conclusions}

{\it Euclid} will be able to generate a sample of $\sim10^3$ strong
gravitational lens events with known background redshifts and
$\sim100$\%\ reliability by selecting $>12L_*$ H$\alpha$ emitting
galaxies at redshifts $z<1.7$; the {\it SKA} will similarly be able to
compile up to $\sim10^5$ lensing events with $\sim100$\%\ reliability and
known source redshifts by selecting $>12M_*$ systems in an H{\sc i}
survey at $z<2$.



\acknowledgments {\boldforreferee The author thanks the
  anonymous referee for helpful suggestions, the Science and
  Technology Facilities Council for support under grant ST/J001597/1,
  and many colleagues in the {\it Euclid} Strong Lensing and Galaxy
  Evolution working groups for stimulating discussions.}



{\it Facilities:} \facility{Euclid}, \facility{SKA}.

\clearpage

\end{document}